\documentstyle[12pt]{article}

\newcommand{\secn}[1]{section~\ref{#1}}

\newcommand{\eq}[1]{Eq.~(\ref{#1})}

\newcommand{\nl}{\nonumber \\}

\def\beq{\begin{equation}}
\def\eeq{\end{equation}}
\def\beqa{\begin{eqnarray}}
\def\eeqa{\end{eqnarray}}

\begin{document}

\rightline{DFTT 14/96}
\rightline{NORDITA 96/25-P}
\rightline{\hfill April, 1996}

\vskip 3cm

\centerline{\Large \bf QCD corrections to the production of a} 
\centerline{\Large \bf heavy quark pair plus a hard photon}
\centerline{\Large \bf in $e^+ e^-$ annihilation}

\vskip 1cm

\centerline{\bf Lorenzo Magnea\footnote{On leave from
Universit\`a di Torino, Italy}}
\centerline{\sl NORDITA}
\centerline{\sl Blegdamsvej 17, DK-2100  Copenhagen \O, Denmark}

\vskip 1cm

\centerline{\bf Ezio Maina\footnote{e-mail: maina@to.infn.it}}
\centerline{\sl Dipartimento di Fisica Teorica, Universit\`a di Torino}
\centerline{\sl Via P. Giuria 1, I-10125 Turin, Italy}
\centerline{\sl and INFN, Sezione di Torino}

\vskip 1cm

\begin{abstract}

We present complete results for the $O(\alpha_s)$ corrections to the
production of a heavy quark pair plus a hard photon, in $e^+ e^-$ annihilation,
including both photon and $Z$ intermediate states.
Virtual corrections are calculated analytically in $4 - 2 \epsilon$ dimensions,
together with the soft approximation to the real emission diagrams.
After cancellation of infrared divergences, real and virtual contributions
are combined numerically to obtain the cross section for 
the production of two heavy quark jets plus a hard photon in next-to-leading 
order. The resulting fully differential cross section can be used to construct
arbitrary distributions with the desired experimental cuts.

\end{abstract}

\newpage

\section{Introduction}
\label{intro}

Out of several millions of $Z$ bosons produced by LEP1, several tens of
thousands decayed into heavy quark pairs. This provides enough statistics
to compare theory and experiment beyond leading order, and possibly to isolate
quark mass effects in the radiative corrections. 
On the other hand, the measured partial widths of the $Z$ to $c$ and $b$ 
quarks, $\Gamma_c$ and $\Gamma_b$, differ by a few standard deviations 
from the predictions of the Standard Model (SM)~\cite{exp_gamma}, and provide 
the most significant discrepancy between high precision data and the 
otherwise spectacularly successfull SM. 
This ensures that all aspects of heavy flavour production will be actively 
studied also in the future.
Finally, quark mass corrections
are going to be extremely important at any of the next-generation $e^+ e^-$
colliders that are currently being planned, where top quark pairs are going
to be abundantly produced, while the top quark mass will be a significant
fraction of the center of mass energy. 
At leading order it has been shown that
quark masses substantially affect three, four and five jet cross sections
\cite{BMM,BM}.
However, NLO calculations for
heavy quark production in $e^+ e^-$ annihilation are available only for 
a two-particle final state~\cite{zer}. 
With the present paper, we begin an analysis of heavy quark production 
processes in $e^+ e^-$ annihilation with more than two particles in the 
final state, at next-to-leading order in QCD.
Here we present results for the production of two heavy quark jets plus 
a hard photon, at $O(\alpha^3 \alpha_s)$. While, as we shall see, this
process is of interest for several reasons, the calculation can also be
considered as a preliminary step towards the complete evaluation of the 
$3$-jet production process at $O(\alpha^2 \alpha_s^2)$, with massive quarks.
We are thus interested in the mass corrections to the
results of Ref.~\cite{kra}, and in perspective to the results of 
Ref.~\cite{ell}.

The importance of studying events with hard photons and hadrons in the final
state, in $e^+ e^-$ collisions, has been emphasized by many authors~\cite{all}.
In particular, it has been argued that photon radiation can be used for a
relatively clean determination of the weak couplings of the radiating 
quark~\cite{mat}. Further, photon production is expected to give signals 
of several new physics processes, including Higgs production~\cite{rid}, 
both in the Standard Model and in its supersymmetric extension, and exotic 
decays of heavy quarks~\cite{ali}. A full understanding of QCD corrections
to the standard production process is however essential in all these cases.
Here we will concentrate on just these corrections, in the case in which the
mass of the quarks is not negligible.

The analysis of hard photon events, and in particular of events with isolated
photons, with which the present work is concerned, is complicated both 
theoretically and experimentally by the necessity to impose cuts to define
the event and to avoid the soft singularities of perturbative calculations.
The difficulties arising were thoroughly discussed, in the case of massless  
quarks, in Refs.~\cite{kun},~\cite{glost} and~\cite{MSZ}. Experimental results
on the production of hard isolated photon at Lep can be found in 
Ref.~\cite{expLep} and references therein.
The case of
massive quarks is somewhat different, and in fact simpler, because the quark
mass regulates the collinear photon-quark and gluon-quark divergences, which
are softened to logarithms of the quark mass; only the infrared divergences
associated with soft emission remain to be dealt with. While in
Ref.~\cite{kun} it was argued that a genuine non-perturbative contribution 
survives even for isolated photons, signalled by the collinear divergence 
associated with photon emission, which must be reabsorbed in a non-perturbative
photon fragmentation function, in the case at hand a large quark mass acts
as a natural cutoff for non-perturbative effects. A partial breakdown of 
perturbation theory is signalled only by the appearance of potentially large 
logarithms of the quark mass, which may eventually be resummed.
It was also argued in Ref.~\cite{qiu} that isolated photon cross sections
suffer from a failure of the usual factorization theorem. However such
a failure takes place only at the boundary of phase space, and does not 
significantly affect our results.

We now proceed to describe how we deal with photon isolation and with the 
cancellation of infrared divergences. Next we will describe briefly the 
steps involved in the calculation of the various contributions to the
cross section. In Sec.~\ref{virt+re} we will deal 
with the virtual corrections, 
which constitute the bulk of the analytic calculations and 
we will briefly describe the real emission contributions.
In Sec.~\ref{res} we will give our results. 

\section{Photon isolation and cancellation of infrared divergences}
\label{iso}

As with all NLO calculations, to obtain a finite answer for a process with 
$p$ partons in the final state, it is necessary to combine the virtual 
one-loop corrections to the production of $p$ partons with the tree-level 
contributions having $p + 1$ partons in the final state, 
integrated over the singular regions of phase space. The exclusive production
of a photon in the final state is not, as a consequence, an infrared safe
process, since diagrams in which the photon is virtual are excluded. It is 
then necessary, both theoretically and experimentally, to introduce a cut
in photon energy, $E_{min}$, to exclude infrared photons. In a hadronic
environment, with hadrons clustered into jets, it is further desirable
to isolate the photon from hadronic jets. This makes for a cleaner
experimental definition of the event, while from a theoretical point of view
it excludes the regions corresponding to collinear emission, in which 
perturbation theory fails or becomes unreliable. This angular isolation 
is the source of most of the ambiguities studied in \cite{kun,glost,MSZ}, since
perfect isolation cannot be achieved either theoretically or experimentally,
and it is necessary to clarify the interplay of the cuts with the jet
reconstruction algorithms. Let us first explain in some detail the
handling of infrared divergences associated with the gluon, then we will
introduce the auxiliary cuts needed for photon isolation.

For the cancellation of infrared divergences, we use here a simplified version
of the ``subtraction method'' described in Ref.~\cite{man}. We have also
tested the ``phase space slicing method'' of Ref.~\cite{gie}, and the results
of the two methods are fully compatible.
Our procedure can be summarized as follows. 
Let $d\sigma^{(3)}_1$ be the differential cross section for the
production of a heavy quark-antiquark pair plus a hard photon, at one loop
in QCD, and let $d\sigma^{(4)}_0$ be the tree-level differential cross section
for the production of the same final state with an extra gluon. Both cross
sections are singular, due to the infrared divergence associated with the 
extra gluon, while in this case there are no collinear divergences, since 
they are cut off by the quark mass. 
In both cases, the singular contributions can be
isolated and computed analytically in $4 - 2\epsilon$ dimensions. 
Let us denote these singular contributions by $d\sigma^{(3)}_{1,soft}$ and 
$d\sigma^{(4)}_{0,eik}$, respectively. Both are proportional to the Born
cross section for the $3$-particle final state, $d\sigma^{(3)}_0$. If we now
integrate $d\sigma^{(4)}_{0,eik}$ over the soft gluon phase space
(defined for example by a cutoff in the quark-gluon invariant mass, or in the
gluon energy in a suitable frame, which we denote here by $\Delta$) 
the result, $d\sigma^{(4 \rightarrow 3 )}_{0}(\Delta)$, will cancel 
the IR singularity of $d\sigma^{(3)}_{1,soft}$, by the Bloch-Nordsieck 
mechanism. We can thus define a finite NLO cross section for the production 
of two heavy quark jets plus a hard photon as
\beq
d\sigma^{(2 jets + \gamma)}_1 = \left[ d\sigma^{(3)}_1 + 
     d\sigma^{(4 \rightarrow 3)}_{0}(\Delta) \right] +
     \left[ \int_R d\sigma^{(4)}_0 - 
      \int_\Delta d\sigma^{(4)}_{0,eik} \right]~~~.
\label{sigjet}
\eeq
Here the first bracket is finite by the Block-Nordsieck mechanism, while the
second one is finite by construction. $R$ is the region in phase space in which
the soft gluon is not experimentally detectable, and it is unrelated with
the parameter $\Delta$ introduced before. The first bracket
can be computed analytically in $4 - 2\epsilon$ dimensions up to the point
where the cancellation of divergences can be explicitely checked.
The resulting finite expression can then be computed numerically.
The second bracket, being finite in four 
dimensions, can be evaluated numerically directly in $d = 4$ using a numerical
integration program. This is important in this case since the exact expression
for $d\sigma^{(4)}_0$ is long and cumbersome. Notice that no approximations 
have been introduced in Eq.~\ref{sigjet}, since exactly the same term was 
added and subtracted.

So far only the cutoff $\Delta$ has been introduced, which defines the border 
of the soft gluon region. Next, we need to restrict the photon phase space
to define an isolated photon cross section, without spoiling the cancellation
outlined above. This we do as follows.
First of all, we introduce energy and transverse momentum cuts to
exclude soft photons and photons lost in the beam pipe. The second of these
cuts is also necessary to avoid the collinear divergence associated with 
collinear photon emission from the massless lepton lines in the initial state.
Thus events with photons having an energy $E_\gamma < E_{min}$ or a 
transverse momentum $p^T_\gamma < p^T_{min}$ in the center of mass frame
are rejected. Next, at the parton level, we impose a cut 
$\theta_{min}$ on the angle
between the photon and either the quark or the antiquark. An event is
rejected if the photon is too close to them. The gluon is treated differently;
soft gluons must be integrated over full phase space to cancel the infrared
divergences of the virtual correction, so we must allow soft gluons to be
emitted in the vicinity of the photon. This is achieved by introducing
an auxiliary infrared cutoff $\Delta_0 < \Delta$ on, say, the quark-gluon 
invariant mass $m_{Qg}$. 
Then for gluons such that $m_{Qg} < \Delta_0$ no isolation 
cut is imposed, whereas gluons with $m_{Qg} > \Delta_0$ are not allowed 
inside the isolation cone of the photon. We remark that treating quarks and
gluons differently is quite natural in the case of massive quarks since these
can in principle be tagged.
At the parton level this defines a cross
section for the production of a hard isolated photon in association
with up to three partons, with a finite amount of partonic energy allowed
inside the isolation cone. 
The events that have passed the cuts are then 
fed through a jet reconstruction algorithm, say Jade, or Durham, where only
the partons, but not the photon, are recombined into jets. Finally, we test if
the isolation of the photon has survived jet reconstruction, using the same 
separation parameter $y_{cut}$ for the definition of the number of jets and for
photon isolation.
It may happen in fact that two soft partons that were outside the photon 
isolation cone are recombined by the jet algorithm into a jet whose axis 
lies inside the cone. 
Such events do not correspond to the experimental definition of an
isolated hard photon, and are again discarded.

Summarizing, our definition of the isolated photon cross section depends
on five phase space cuts, $E_{min}$, $ p^T_{min}$, $\theta_{min}$, $\Delta$
and $\Delta_0$. Consistency demands that our results be independent of
the definition of the soft gluon region, so they should be stable against
variations of the infrared cutoffs $\Delta$ and $\Delta_0$.
As explained in more detail in \secn{res}, we have verified that this 
is indeed the case.

\section{Outline of the calculation}
\label{virt+re}

The handling of virtual corrections is straightforward, if lengthy.
Since the masses of the heavy quarks are comparatively well known, we
choose on-shell renormalization for the quark mass and wave function.
As a consequence, quark self-energies on the external legs do not
contribute, except for the fact that an infrared divergence is introduced
in the quark wavefunction renormalization, which will cancel when combined
with real gluon emission, as in QED. 
The strong coupling is not renormalized,
since the corrections computed here are leading order in $\alpha_s$. 

To fix our notations and normalizations, let us start by displaying the 
structure of the relevant matrix element. We consider the reaction
\beq
\label{reaction}
e^-(p) + e^+(p^\prime) \rightarrow Q(p_1) + \bar{Q}(p_2) + \gamma(p_3)~~~,
\eeq
where $Q$ is a quark of mass $m >> \Lambda_{QCD}$, {\it i. e.} a $c$, $b$
or $t$ quark. Including photon emission from the initial state leptons,
the matrix element for the reaction in \eq{reaction} can be written as
\beqa
\label{mel}
M(p_1,p_2,p_3;s,\mu) & = & \frac{1}{s} \left( L_\mu^{(0)}(\gamma) 
H_{(1)}^\mu (\gamma) + L_\mu^{(1)}(\gamma) H_{(0)}^\mu (\gamma) \right) \\
& + & \frac{1}{s - M_Z^2 + {\rm i} \Gamma_Z M_Z} \left(
L_\mu^{(0)}(Z) H_{(1)}^\mu (Z) + L_\mu^{(1)}(Z) H_{(0)}^\mu (Z) 
\right)~~.  \nonumber
\eeqa
Here $L_\mu^{(i)}(V)$ is the leptonic vertex for the production of
vector boson $V$, including real photon radiation if $i = 1$,
while $H_\mu^{(i)}(V)$ is the corresponding hadronic vertex.
Specifically,
\beqa
\label{lep0}
L_\mu^{(0)}(\gamma) & = & - {\rm i}~ e~\bar{u}_e(p) 
\gamma_\mu v_e(p^\prime) \nl
L_\mu^{(0)}(Z) & = & - {\rm i}~ 
\frac{e}{2 \sin \theta_W \cos \theta_W}~\bar{u}_e(p) 
\gamma_\mu \left(c_V^e - c_A^e \gamma_5 \right) v_e(p^\prime)~~~,
\eeqa
where the electron couplings to the $Z$ boson are given by
\beq
\label{ecoup}
c_V^e = - \frac{1}{2} + 2 \sin^2 \theta_W~~~,~~~~~~~~~c_A^e = - \frac{1}{2}~~~.
\eeq
Initial state photon radiation is contained in $L_\mu^{(1)}(V)$, which is the
sum of the two diagrams obtained by inserting a photon on either of the
leptonic legs in \eq{lep0}.
Similarly we can write
\beq
\label{hadver}
H_\mu^{(i)}(V) = H_\mu^{(i,0)}(V) + \alpha_s H_\mu^{(i,1)}(V)~~~.
\eeq
Here the tree level contributions $H_\mu^{(0,0)}(\gamma)$ and
$H_\mu^{(0,0)}(Z)$ are obtained from \eq{lep0} by replacing the electron
couplings to the various currents with the corresponding quark couplings.
Final-state tree-level photon emission is contained in $H_\mu^{(1,0)}(V)$,
which is again the sum of two diagrams for each gauge boson.
$H_\mu^{(0,1)}(V)$ is the QCD radiative correction to $Q \bar{Q}$
production, and, since we are using on-shell renormalization for the quark 
lines, it includes only the vertex correction diagram. 
Finally, the bulk of the computation is devoted to $H_\mu^{(1,1)}(V)$, the 
QCD correction to photon emission from the final state quark lines.
Not counting quark self-energies on external legs, 
this correction comprises eight diagrams, two of which are of ``box'' type.

The first step in the analysis of the hadronic vertex is a 
Passarino-Veltman decomposition~\cite{gian} of the tensor integrals
contained in $H_\mu^{(i,1)}(V)$, reducing them to combinations of scalar 
integrals.
This decomposition can be performed automatically using a symbolic 
manipulation program, in our case FORM~\cite{verma}. 
To make sure of the result,
we checked it using a corrected and adapted version 
of a FORM program by J. Vermaseren based on the methods discussed in 
Ref.~\cite{Vermaseren}.

Having separated the tensor structure of the hadronic vertex from the scalar
integrals, the next step is to write down the matrix element as a linear
combination of a fundamental set of Lorentz structures involving spinors,
polarization vectors, Dirac matrices and external momenta, with scalar
coefficients. This is the method of standard matrix elements,
described in detail in Ref.~\cite{denner}.

In the present case, after using momentum and current conservation, as well
as the mass-shell conditions, the hadronic vertices $H_\mu^{(0,1)}(V)$ can be
expressed as combinations of $4$ standard matrix elements. The more 
complicated vertices $H_\mu^{(1,1)}(V)$, which carry two vector indices and
depend on three momenta, can be expressed as linear combinations of
$40$ standard matrix elements. The coefficients of these linear combinations 
are deduced from the Passarino-Veltman procedure.
In the algebraic manipulations which reduce $H_\mu^{(1,i)}(V)$ ($i = 0, 1$) 
to standard matrix elements, $\gamma_5$ has
been assumed to anticommute with all $\gamma$ matrices. This is
consistent in the present calculation, which contains no anomalous 
contributions. In particular, it can be seen that a na\"{\i}ve implementation
of the method of standard matrix elements corresponds to the treatment
of $\gamma_5$ suggested in Ref.~\cite{dirk}.
The resulting decomposition into standard matrix elements multiplied by 
scalar form factors was automatically translated into FORTRAN
code, to be readily implemented in a numerical evaluation of the
square of the matrix element. 

The final step is the evaluation of the scalar integrals, which is
done analytically, using dimensional regularization both for 
ultraviolet and infrared divergences. The non-trivial integrals
are four ``triangle'' integrals and one ``box'' integral, which
can all be evaluated using standard methods. 
It should be noticed that for the present 
purposes it is important to keep the imaginary parts of the scalar
integrals, since the $Z$ propagator has an imaginary part, and
there will be real contributions to the square of the matrix element
arising from the interference of the $Z$ width with the imaginary parts
of the scalar integrals.

After renormalization, virtual corrections still contain
infrared divergences, which must cancel against real soft gluon emission.
The infared divergent part of the one-loop correction must be given
by the well-known universal soft factor, multiplied by the tree-level
contribution. This provides a first non-trivial analytic check of the 
algebraic manipulations performed on the matrix element.
We have further tested the matrix element, computed numerically using our 
program, by verifying gauge invariance with respect to the external photon,
and we have also tested the conservation of the vector current on the 
hadronic vertex in the case in which this conservation
is not used in the construction of the standard matrix elements.
A similar test for the axial current is non-trivial since we are working with
massive quarks, so that the axial current is not conserved, but rather related
by a Ward identity to pair production by a pseudoscalar source.

Real gluon emission corresponds to the process
\beq
e^-(p) + e^+(p^\prime) \rightarrow Q(p_1) + \bar{Q}(p_2) + \gamma(p_3) +
g(p_4)~~~,
\label{alreaction}
\eeq
which admits a decomposition analogous to \eq{mel}. The amplitude for
this process is known, and for its evaluation we have used the codes 
described in Ref.~\cite{BMM}. In the soft gluon region, the
amplitude factorizes as usual into the product of an eikonal factor times
the tree-level amplitude for the process in \eq{reaction}. The resulting 
infrared divergent contribution to the cross section can easily be
calculated analytically, and it cancels the corresponding divergence due
to virtual gluon exchange, according to \eq{sigjet}.

\section{Results}
\label{res}

In this paper we concentrate on the production of a photon in association with 
$b$ quarks at LEP1. 
We take $m_b = 4.7$ GeV, $\alpha_s = \alpha_s(M_Z) = 0.122$, 
$\alpha = 1/128.07$, $\sin^2 \theta_W = 0.2247$, $M_Z = 91.187$ GeV, 
and $\Gamma_Z = 2.49$ GeV.

In Fig.1 we present the cross sections $\sigma_n = \sigma(\gamma + n~jet)$
for the production of a photon 
plus $n$ jets, with $n = 1,\ 2,\ 3$, at leading order (LO) and including
${\cal O}(\alpha_s )$ corrections (NLO). The center--of--mass energy is set to
$\sqrt{s} = M_Z$, and the cross sections are given as functions of $y_{cut}$,
including both photon and $Z$ intermediate
states, as well as the contribution of hard photon emission from 
the $e^+e^-$ pair.
We neglect the effect of soft multiple initial state radiation.
Events are accepted if the photon energy and transverse momentum are in the
range 
\beq
E_\gamma > 10 \hspace{.2 cm} {\rm GeV} \hspace{1 cm},~~~~~~
p^T_\gamma > 5 \hspace{.2 cm} {\rm GeV}~~~.
\label{cuts}
\eeq
A minimum angle $\theta_{min} = 15^\circ$ is required between the photon and
both quarks. For the real contribution, if the gluon is such that 
$m^2_{Qg} > \Delta^2_0 = m_b^2 + 0.1 = 22.19$ GeV$^2$, the same minimum angle 
is required between the photon and the gluon. We have tested that our results
are not sensitive to the particular choice of $\Delta_0$. In particular,
varying the ratio $(\Delta^2_0 - m_b^2)/m_b^2$ by two orders of magnitude
affects typical observables at the level of $ 1-2 \% $.

Jets are reconstructed using the invariant mass squared of the parton pairs
\beq
y_{ij}= \frac{(p_i+p_j)^2}{s}.
\eeq
However, our code is flexible and other jet reconstruction algorithms can be
readily adopted.
As we mentioned in \secn{iso}, the single parameter $y_{cut}$ is used both to 
define the number of jets in the event and to test the photon isolation 
from the jets, after reconstruction.
With the present set of cuts the contribution of photon emission from the
lepton pair amounts to about 5\% of the total cross section.

In Fig. 2 we show the fractional change 
$R_i = (\sigma_i^{NLO}-\sigma_i^{LO})/\sigma_i^{LO}$ of NLO to LO 
cross section for $\sigma_1$ and $\sigma_2$. 
Finally, in Figs. 3, 4 and 5 we show a few examples of 
observable distributions 
that can be studied  using the present calculation. 
Fig. 3 shows the photon energy distribution, Fig. 4 the transverse momentum
one, while Fig. 5 deals with the cosine of the angle with respect to 
the $e^-$ direction. All distributions are given both at LO and at NLO,
are obtained for $y_{cut} = 10^{-2}$, and are summed over the number of jets 
produced in association with the photon.

Notice that the cross section $\sigma_1$ is hardly modified 
by QCD corrections. In the $y_{cut}$ range we have examined $\sigma_1$ 
is always very small, at most of the order of 10\%.
The cross section $\sigma_3$ at ${\cal O} (\alpha_s)$
is completely determined by tree--level real emission diagrams and shows the
usual steep rise for small values of the jet--jet separation parameter, which
will eventually be tamed by the inclusion of ${\cal O} (\alpha_s^2)$
corrections. 
The most interesting result is the strong modification of the rate
for two jets plus photon production. At large values of 
$y_{cut}$, $\sigma_2$ increases
with respect to the LO result due to the contribution of three parton 
events in which two of the partons are unresolved. 
The growth of the cross section at smaller value of $y_{cut}$ is slower at 
NLO until, at about $y_{cut} = 2 \times 10^{-2}$, $\sigma_2$
begins to decrease quite sharply. At  $y_{cut} = 1 \times
10^{-2}$, $\sigma_2 \approx \sigma_3$.

The size of NLO radiative corrections is illustrated in Fig.2. The corrections 
are relatively small for the one--jet case and positive. They are negligible
at $y_{cut} = 1 \times 10^{-1}$ and increase for decreasing $y_{cut}$, reaching
30\% at $y_{cut} = 1 \times 10^{-2}$. For the two--jet case they are negative
for $y_{cut} < 5 \times 10^{-2}$ and positive for larger separations.
The corrections are about --60\% at $y_{cut} = 1 \times 10^{-2}$ and
+30\% at $y_{cut} = 1 \times 10^{-1}$, 
much like in the massless case, as shown 
for example in Ref.~\cite{glost}. The corrections to $\sigma_1$
in the massive quark case have the opposite sign with respect to the massless
quark result and are much smaller in magnitude.

The shapes of the differential distributions are not strongly affected by 
${\cal O} (\alpha_s)$ corrections. The only modification of some interest 
is the kink at $E_\gamma \approx 22.5$ GeV in the photon energy spectrum. 
This kink can be traced back to the behaviour of the box diagrams, 
whose contribution is large and positive. Box diagrams correspond
to a rescattering of the quark and antiquark after photon emission has taken
place. As a consequence of this rescattering, high energy photons
are less suppressed at one loop than at tree level, where the corresponding
matrix element must contain a far off shell propagator.
The differential $E_\gamma$ distribution of all remaining diagrams
is smooth at all energies. These latter diagrams cancel to a large 
extent, resulting in a relatively small negative contribution. 
The presence and the location of the kink are insensitive to $E_{min}$, 
$ p^T_{min}$, $\theta_{min}$ and to the quark mass. In fact we have checked 
that the same behaviour is reproduced for a $c$--quark
with $m_c = 1.7$ GeV. On the other hand, varying the center--of--mass 
energy the location $E_k$ of the
kink does change. In the range 80 GeV $<\ \sqrt{s}\ <$ 120 GeV we find
$E_k \approx \sqrt{s}/4$. This effect is present even if in pure QED, 
namely if we only consider the contribution of the photon intermediate state
neglecting the diagrams with a virtual $Z$.
Given the limited statistics which can be expected, we are skeptical 
about the possibility of measuring the energy spectrum with enough accuracy 
to test for this kind of behaviour. 

Notice finally that all the results are affected by an unphysical logarithmic
dependence on the scale $\mu$ chosen for the running QCD coupling, as with
all calculations that are of leading order in $\alpha_s$.

\section{Conclusions}
 
We have presented complete results for the $O(\alpha_s)$ corrections to the
production of a heavy quark pair plus a hard photon, in $e^+ e^-$ annihilation,
including both photon and $Z$ intermediate states. Our results are implemented
in a computer code that evaluates numerically the relevant matrix elements
and combines them using a jet reconstruction algorithm to give the cross 
sections for hard photon production in association with $n$ jets, with 
$n = 1$, $2$ and $3$. The resulting fully differential cross sections can be 
used to construct arbitrary distributions, with the desired experimental cuts.

We have discussed in detail the production of a photon in association with 
$b$ quarks at LEP1, and we have given a few examples of relevant distributions 
in that case. We observe that the shape of the cross section for the production
of $2$ jets plus a resolved photon is significantly affected by NLO
corrections, as may be expected, since the small-$y_{\rm cut}$ growth of
the tree level result must be tamed by radiative corrections. The shapes of 
the other differential distributions is not significantly affected by radiative
corrections. However it is possible to see the effects of the correlations
between partons induced by gluon exchange, albeit at a level probably
inaccessible experimentally. The overall normalization of the distributions
is shifted downwards by QCD corrections, as may be expected, since virtual 
corrections are negative, while the inclusion of gluon radiation restricts 
the available phase space for hard photon emission, so that comparatively 
more photon events are excluded by the cuts.

The present work can be readily extended to other heavy quarks and higher
energies, notably to the case of photon production in association with top 
pairs at future linear colliders. Work on these applications, as well as 
on a detailed comparison with the massless quark case, is in progress.

\vskip 2cm

{\large {\bf Acknowledgements}}
\vskip 0.2cm

We would like to thank Alessandro Ballestrero for many helpful discussions
and for checking several parts of the computer code.
L. M. would like to thank Dirk Kreimer for a helpful conversation, as well
as Lance Dixon for his kind invitation to the Aspen Institute in the summer
of 1995, where several discussions concerning the present work took place.

\newpage
{\large {\bf Figure Captions}}
\vskip 0.2cm
\begin{description} 
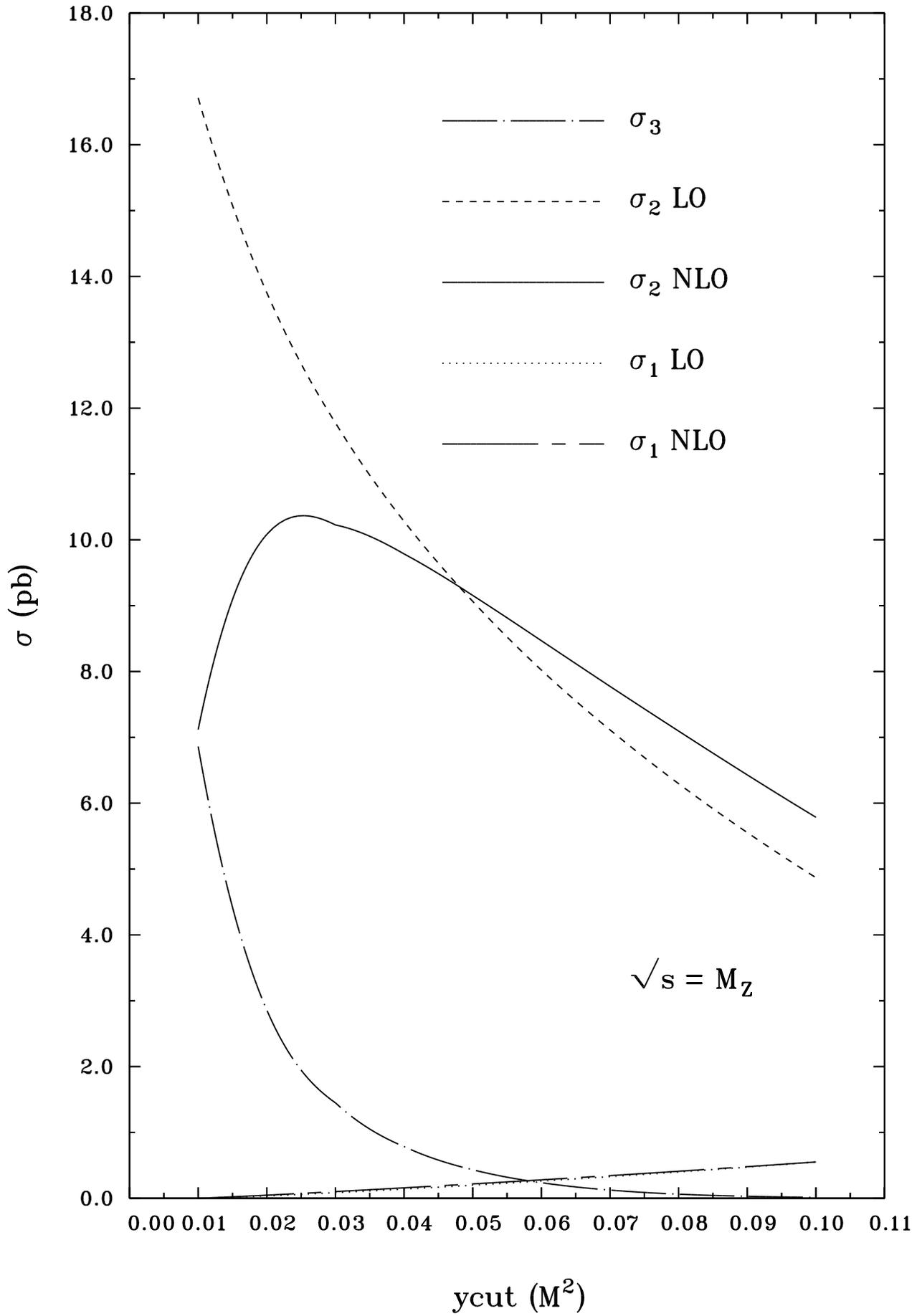
\item[Fig.1] 
Cross sections $\sigma_n = \sigma(\gamma + n~jet)$
for the production of a photon 
plus $n$ jets, with $n = 1,\ 2,\ 3$, at LO and at
NLO in QCD as functions of $y_{cut}$.
Hard photon emission from the $e^+e^-$ pair is included. The jet reconstruction
procedure and the set of cuts which ensure photon isolation are described 
in the text.

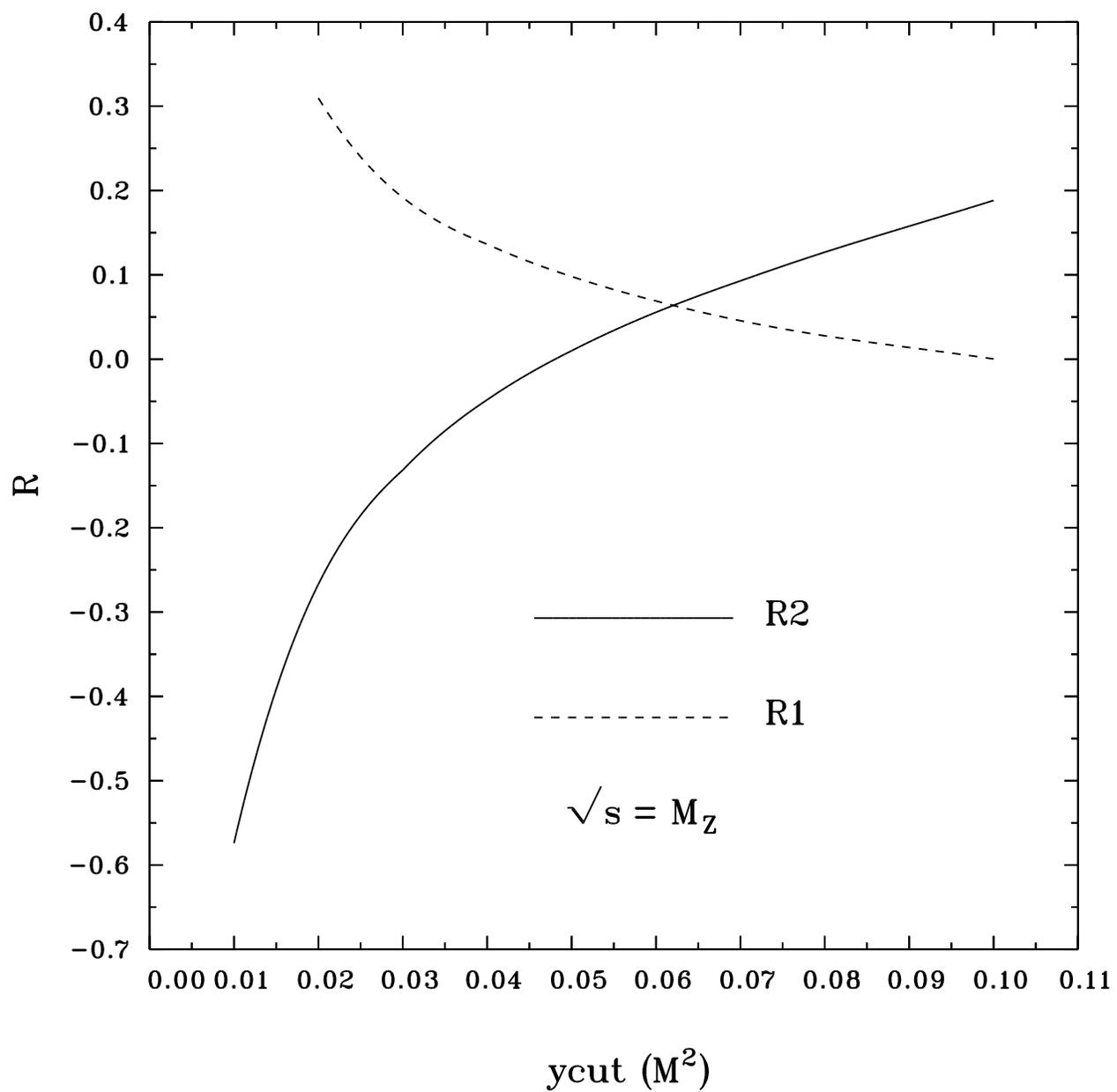
\item[Fig.2] 
Fractional difference $(\sigma_i^{NLO}-\sigma_i^{LO})/\sigma_i^{LO}$ for the
production of a photon plus one and two jets as functions of $y_{cut}$.

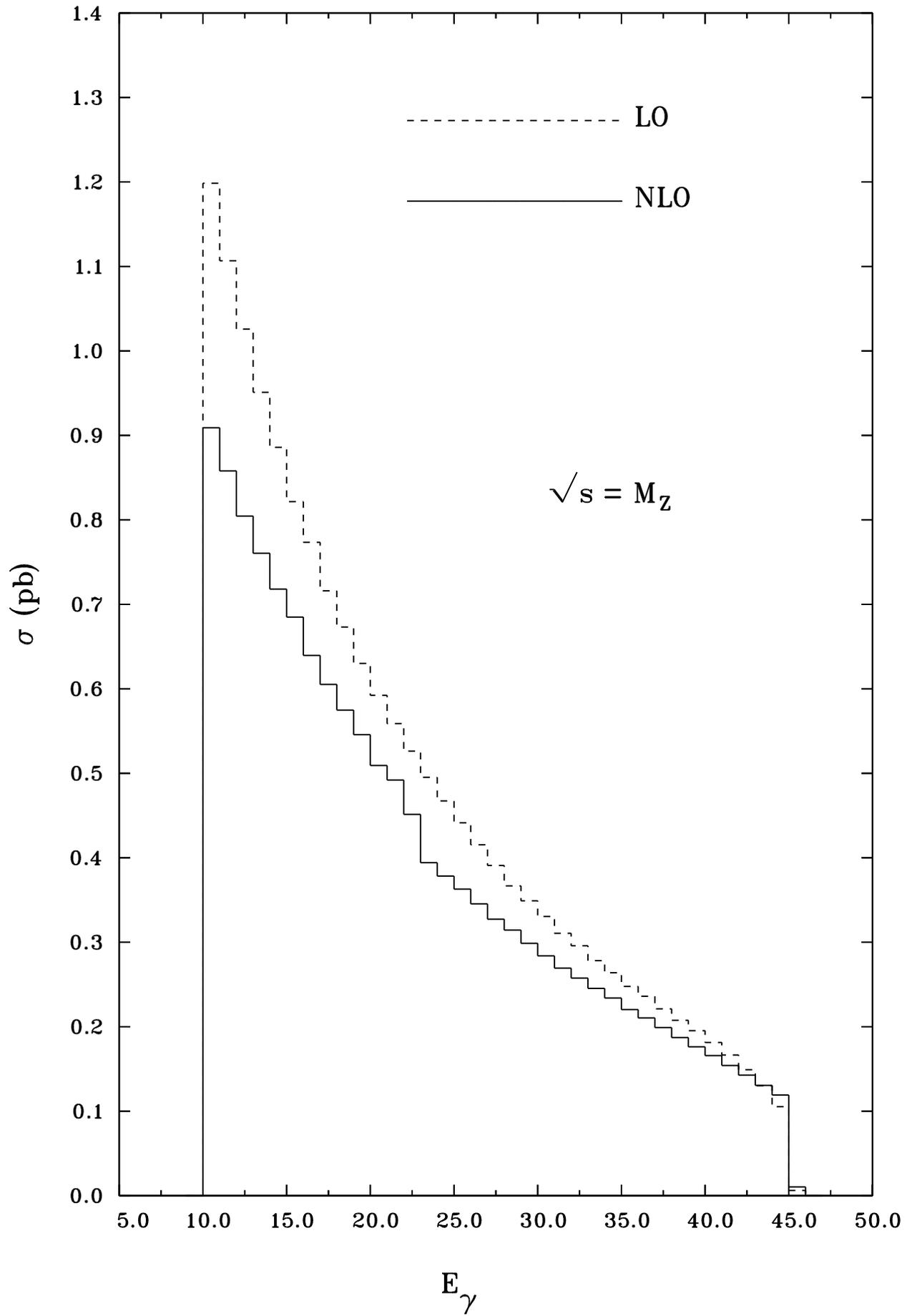
\item[Fig.3] 
Photon energy spectrum at LO and at NLO in QCD for $y_{cut}= 1\times 10^{-2}$.
The separation parameter is $y_{ij}= (p_i+p_j)^2/s$.

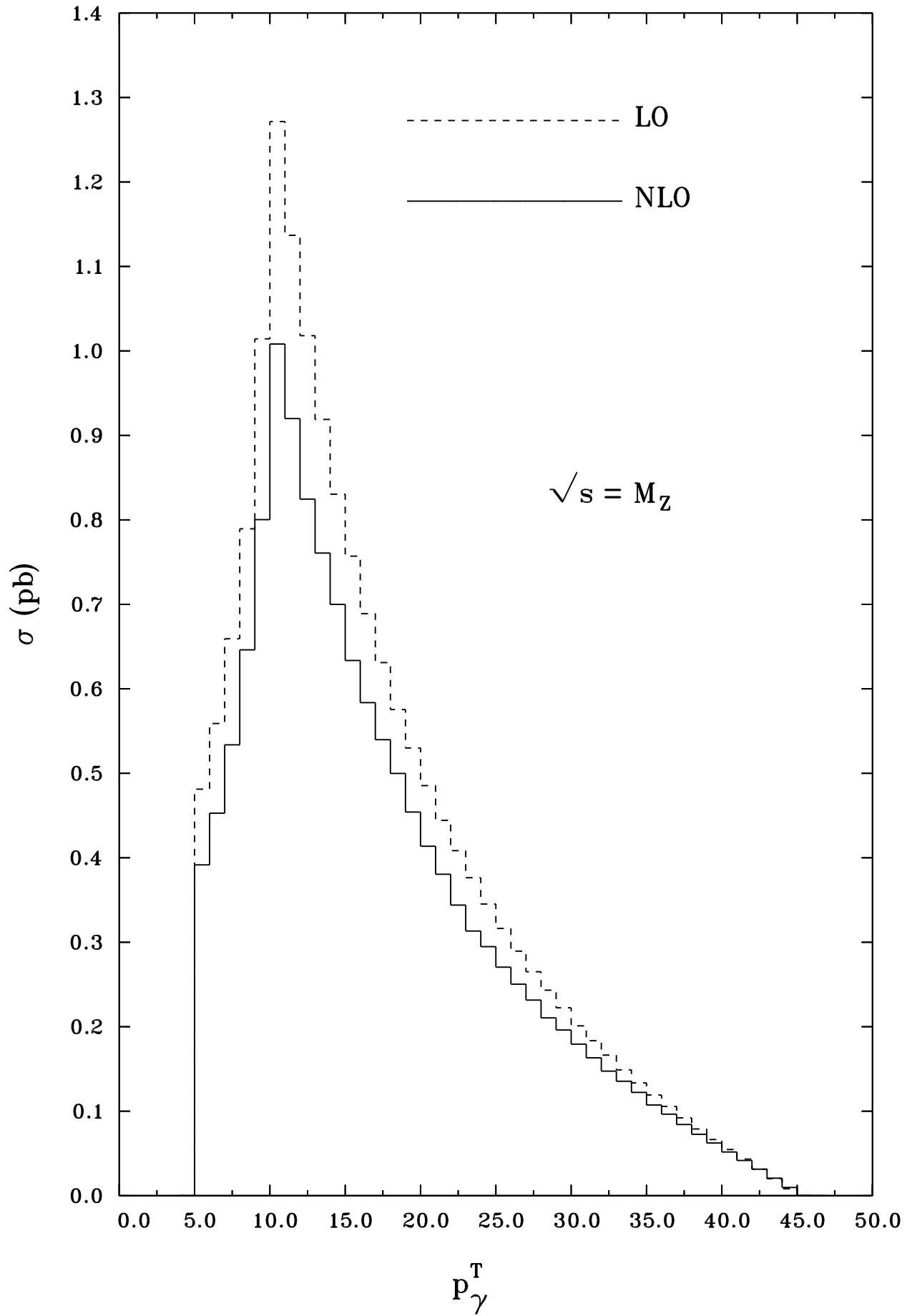
\item[Fig.4] 
Tranverse momentum distribution of the photon at LO and at
NLO in QCD for $y_{cut}= 1\times 10^{-2}$.

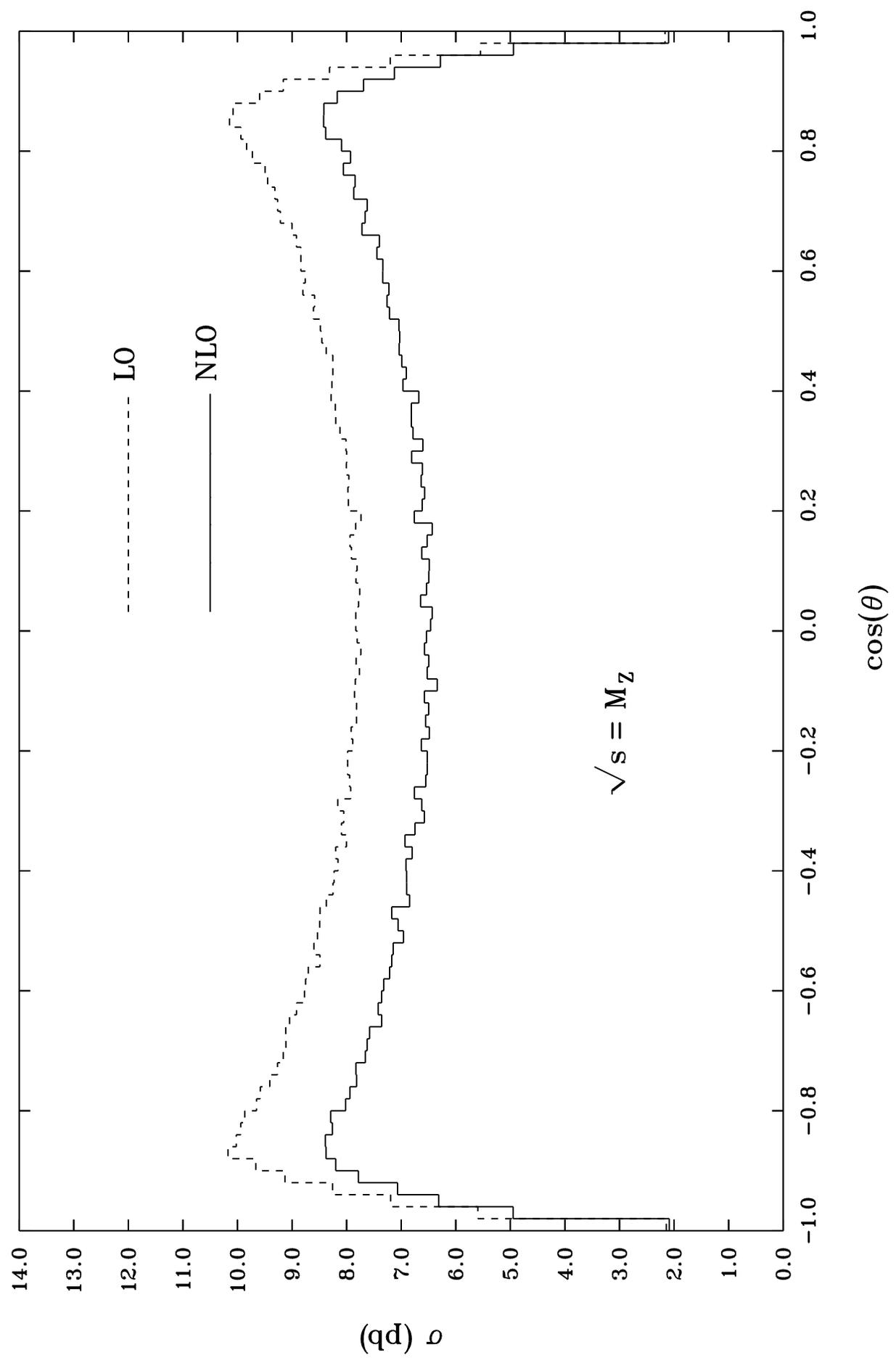
\item[Fig.5] 
Distribution of the cosine of the angle between the photon and the 
incoming electron at LO and at NLO in QCD for $y_{cut}= 1\times 10^{-2}$.

\end{description} 

\end{document}